\documentclass[letterpaper, 10 pt, conference]{ieeeconf}  % Comment this line out if you need a4paper
\usepackage{graphicx}
\usepackage{amsfonts}
\usepackage{algorithm2e}

\usepackage{amsthm}
\usepackage{graphics}
\usepackage{amssymb,amsmath,amsthm}
\newtheorem{definition}{Definition}
\theoremstyle{definition}

\SetAlFnt{\normalsize}
\IEEEoverridecommandlockouts                              % This command is only needed if 
\title{\LARGE \bf
Relationship Explainable Multi-objective Reinforcement Learning with Semantic Explainability Generation *
}

\author{Huixin Zhan$^{1}$ and Yongcan Cao$^{1}$% <-this % stops a space
\thanks{$^{1}$The authors are with the Department of Electrical and Computer Engineering,
        University of Texas, San Antonio, TX, 78249.
        {\tt\small \{huixin.zhan,yongcan.cao\}@utsa.edu}}%
}

\begin{document}

\maketitle
\thispagestyle{empty}
\pagestyle{empty}

%%%%%%%%%%%%%%%%%%%%%%%%%%%%%%%%%%%%%%%%%%%%%%%%%%%%%%%%%%%%%%%%%%%%%%%%%%%%%%%%
\begin{abstract}

Solving multi-objective optimization problems is important in various applications where users are interested in
obtaining optimal policies subject to multiple, yet often conflicting objectives. A typical approach to obtain optimal policies is to first construct a loss function that is based on the scalarization of individual objectives, and then find the optimal policy that minimizes the loss. However, optimizing the scalarized (and weighted) loss does not necessarily provide guarantee of high performance on each possibly conflicting objective because it is challenging to assign the right weights without knowing the relationship among these objectives. Moreover, the effectiveness of these gradient descent algorithms is limited by the agent's ability to explain their decisions and actions to human users. The purpose of this study is two-fold. First, we propose a vector value function based multi-objective reinforcement learning (V2f-MORL) approach that seeks to quantify the inter-objective relationship via reinforcement learning (RL) when the impact of one objective on others is unknown a prior. In particular, we construct one actor and multiple critics that can co-learn the policy and inter-objective relationship matrix (IORM), quantifying the impact of objectives on each other, in an iterative way. Second, we provide a semantic representation that can uncover the trade-off of decision policies made by users to reconcile conflicting objectives based on the proposed V2f-MORL approach for the explainability of the generated behaviors subject to given optimization objectives. We demonstrate the effectiveness of the proposed approach via a MuJoCo based robotics case study.

\end{abstract}

%%%%%%%%%%%%%%%%%%%%%%%%%%%%%%%%%%%%%%%%%%%%%%%%%%%%%%%%%%%%%%%%%%%%%%%%%%%%%%%%
\section{INTRODUCTION}

In recent years, the application of RL in tasks with high-dimensional sensory inputs has shown the potential of creating artificial agents that can learn to accomplish a number of challenging tasks, including the Atari games~\cite{mnih2015human,guo2014deep,schaul2015prioritized,wang2016dueling,van2016deep,oh2015action,nair2015massively}, self-driving cars~\cite{pan2017virtual}, and Go~\cite{maddison2014move,silver2016mastering,silver2018general}. However, the approaches developed therein mainly focus on finding a single usable strategy, without considering the trade-off among potential alternatives that can increase one objective's value at the cost of another.

In the multi-objective setting, the completion of a task requires the simultaneous satisfaction of multiple objectives such as balancing the power consumption and performance in Web servers~\cite{tesauro2008managing}. Such problems can be modeled as multi-objective Markov decision processes (MOMDPs) and solved by some existing multi-objective reinforcement learning (MORL) algorithms~\cite{tesauro2008managing,vamplew2011empirical,roijers2013survey}. However, solutions obtained via these approaches can hardly balance the possibly conflicting objectives to achieve satisfactory performance on all objectives.

Recently, several interesting MORL approaches have been developed. The author in~\cite{nguyen2018multi} proposed the use of both linear weighted sum and nonlinear thresholded lexicographic ordering methods to develop a multi-objective deep RL framework that includes both single- and multi-policy strategies. The author in~\cite{tajmajer2018modular} proposed an architecture in which separated deep Q-networks (DQNs) are used to control the agent's behavior with respect to particular objectives. Then, each DQN has an additional decision value output that acts as a dynamic weight used while summing up Q-values. The authors in~\cite{vamplew2017softmax} used softmax-epsilon selection based on a nonlinear action-selection operator. The agents incorporate an action-selection function that is defined as an ordering over these Q-values. In summary, most of the algorithms are based on the scalarization method to transform the multi-objective problem into a single objective one. The scalarization can be nonlinear or linear~\cite{nguyen2018multi,tajmajer2018modular,vamplew2017softmax,van2013scalarized}. Other advanced methods include, e.g., the convex hull~\cite{roijers2015computing}, the varying parameters approaches~\cite{liu2015multiobjective}, the constraint method~\cite{konak2006multi}, the sequential method~\cite{nakayama2009sequential}, and the max-min method~\cite{lin2005min}.

The authors in~\cite{sener2018multi} proposed an upper bound for the multi-objective loss and proved that optimizing this upper bound via gradient-based multi-objective optimization yields a Pareto optimal solution. The Frank-Wolfe solver is used to find a minimum-norm point in the convex hull of the set of input points. This work provides a new perspective for balancing objectives when the values for all objectives are considered as the min-norm points in the convex hull. This work showed success in large scale multi-label learning tasks. However, it is unclear if the method can be extended to complex continuous space planning tasks.

When the values for all {possibly conflicting objectives} are considered as a vector and balancing them is required, it is critical to train a policy and {explain} why a particular behavior is generated is the topic under study. To address the critical issue, this paper focuses on proposing an explainable V2f-MORL approach. The proposed research has three main contributions. {First}, we propose an approximate optimistic linear support algorithm (AOLS), which allows the quantification of inter-objective relationship using the inter-objective relationship matrix (IORM). {Second}, instead of using scalarized Q-value and the action selection approach based on the priority objective value, the proposed method supports vectorized objective state values. In particular, we propose the creation of multi-objective value functions that can be used sequentially in the training of the critics to update the objective state values and the training of the actor to update the control policy. {Third}, our method is applicable in high-dimensional continuous action spaces with an explainable planning via natural language representation. To our best knowledge, this is the first time that actor critics with quantifiable inter-objective relationship are developed to solve MORL with semantic representation. We also show via one MuJoCo example that the proposed method outperforms the existing single objective optimization methods.

\section{PRELIMINARIES}
\subsection{Multi-Objective Value Function}
For a control policy $\pi$, we here propose the construction of a set of \textit{multi-objective value functions $\mathcal{Y}_{i}^{\pi},~i=1,\cdots,I$} via our defined IORM $W$ as $\mathcal{Y}^\pi=[\mathcal{Y}_{1}^{\pi}, \cdots, \mathcal{Y}_{I}^{\pi}]\in\mathbb{R}^I = W \mathbf{V}^\pi$, where $W\in\mathcal{R}^{I\times I}$ is the IORM, $I$ is the number of objectives, $\mathbf{V}^\pi=[V_1^\pi,\cdots,V_I^\pi]^T$ with $V_i^\pi$ representing the state value for the $i$th objective subject to the control policy $\pi$. The $i$th row of $W$, given by $[w_{i1},\cdots,w_{iI}]$, characterizes how other objective values, $\{V_j^\pi|j\neq i\}$, impact the $i$th objective value $V_i^\pi$ under the policy $\pi$. %multi-objective value functions having $I$ objectives: $\mathcal{Y}_{1}^{\pi}, \mathcal{Y}_{2}^{\pi},..., \mathcal{Y}_{I}^{\pi}$, where $\mathcal{Y}_{i}^{\pi}$ is the expected sum of the values characterizing one specific state value and it's additive coordinates based on other correlated objectives. 
We here provide a formal definition of the multi-objective value function under the policy $\pi$.

\begin{definition}\label{definition2}
Each multi-objective value function is a cumulative sum of objective state values with additive specific impact elements of the form given by
\begin{equation}\label{eq:yipi}
{\mathcal{Y}_{i}^{\pi}\left[i_k\right]}
=\sum_{j=1}^I w_{ij}[i_k]V_j^\pi[i_k],\quad i=1,\cdots,I,
\end{equation}
where $w_{ij}$ is the weight quantifying the impact of $V^\pi_j$ on $V^\pi_i$, $i_k$ is the time step number, and $k$ is the number of sequence.
\end{definition}

%In Definition \ref{definition2}, a scalar $\mathcal{Y}_{i}^{\pi}$ is an inner product of the objective state value vector $V_{\cdot}^{\pi}$ and the basis vector $w_{i\cdot}$. %$I^{th}$ basis includes $n$ impact elements about $n$ vectorized objective state values in~\eqref{eq:yipi}. %$I^{th}$ objective state values are fitted sequencially in the $ik^{th}$ time steps which is showed in Algorithm 2 Line 1. More detailed calculation for the vectorized multi-objective values are shown in Section III.D.

\subsection{Inter-objective Relationship Matrix (IORM)}
Based on Definition~\ref{definition2}, we define the \textit{inter-objective relationship matrix (IORM)} as
\begin{equation}\label{eq:CM}
W= \left [ w_{ij} \right ]\in\mathbb{R}^{I\times I}.
\end{equation} 
%where $w_{ij}$ is defined as the impact of the $i^{th}$ objective on the $j^{th}$ objective. 
Because the impact of one objective on another objective is unknown \textit{a priori}, an IORM can be assigned an initial value but it needs to be updated based on the input observation, denoted as $\mathcal{X}$, and a collection of objective value spaces, denoted as $\left \{ \mathcal{Y}^{i} \right \}_{i\in \left | I \right |}$, for each objective value $y^{i}$ with $m$ input/output examples denoted as $\left(x^{i}_{1},y^{i}_{1}\right),....,\left(x^{i}_{m},y^{i}_{m}\right)$. In particular, IORM will be updated via numerous batches. During each batch,  $I\times k$ time steps will be divided into $I$ sequences. In the $i_k$th time step, $m$ examples are trained to fit $y^i$, which is then used to update the objective state values $V_i$. $\left\{V_j^\pi\left[i_k\right]\right\}_{j\in \left | I \right |}$ is then used to update the $i$th row of the relationship matrix $W$. More detailed description of such an update process will be provided in Section~\ref{sec:PM}.

%The introduction of convex converage set is to quantify the relationship between different policies subject to a number of objectives because one policy can yield good performance for one objective while poor performance for another objective. For instance, for objective $i_{1}$, solution $\theta$ is better when $\mathcal{\hat{L}}^{i_{1}}\left ( \theta_{V_{i_{1}}}, \theta_{V_{j_{1}}}\right )< \mathcal{\hat{L}}^{i_{1}}\left ( \overline{\theta}_{V_{i_{1}}}, \overline{\theta}_{V_{j_{1}}}\right )$, while for objective $i_{2}$, solution $\overline{\theta}$ is better when $\mathcal{\hat{L}}^{i_{2}}\left ( \theta_{V_{i_{2}}}, \theta_{V_{j_{2}}}\right )>  \mathcal{\hat{L}}^{i_{2}}\left ( \overline{\theta}_{V_{i_{2}}}, \overline{\theta}_{V_{j_{2}}}\right )$. $CCS$ defines the set of vector objective state values that the optimal value must reside in because every other vector objective state value not in the set will not be the best choice since there exists at least one vector in $CCS$ that is not smaller than it. 

As a consequence, the vector value function $\mathcal{Y}^{\pi}=[\mathcal{Y}_{i}^{\pi}]\in\mathbb{R}^{I\times I}$, where $\mathcal{Y}_{i}^{\pi}$ is defined in \eqref{eq:yipi}, can be updated via
\begin{equation}\label{eq:movf}
\begin{aligned}
\begin{bmatrix}
\mathcal{Y}_{1}^{\pi}\left[k\right]
\\
%\mathcal{Y}_{2}^{\pi}\left[k\right]
%\\
\vdots
\\
{\mathcal{Y}_{n}^{\pi}\left[k\right]}
\end{bmatrix}
=W[k]\begin{bmatrix}
V_{1}^{\pi}\left[k\right] %\\ V_{2}^{\pi}\left[k\right]
\\
\vdots
\\
{V_{n}^{\pi}\left[k\right] }
\end{bmatrix},
\end{aligned}
\end{equation}
where $W[k]$ is the updated IORM at the $k$th time step.

\section{PROPOSED METHOD}\label{sec:PM}
\subsection{Multi-objective Decision Making}
In this paper, we consider the problem when multiple objectives $O_i,~i=1,\cdots,I,$ need to be optimized for a given mission, where $I$ denotes the number of objectives. For example, in robotic locomotion, maximizing forward velocity but minimizing joint torque and impact with the ground, result in a very large number of options to consider. We use $\mathbf{V}^\pi \in \mathbb{R}^{d\times1} =  [V_1^\pi,\cdots,V_I^\pi]^T$, where $d \geq 2$, to represent the vector value function for $O_i,~i=1,\cdots,I,$ subject to the control policy $\pi$. A typical approach to optimize objectives $O_i,~i=1,\cdots,I,$ is to construct a scalarized value function of the form $V_{w}^{\pi} \in \mathbb{R}^1 =w\mathbf{V}^\pi$, where $w=[w_1,\cdots, w_I]$ satisfying $w\textbf{1}=1$, $\textbf{1}$ is an all-one column vector, and weight $w_i$ specifies how much each objective contributes to the scalarized objective. A more general form of the scalarized value function is given by $V_{w}^{\pi}=f\left(w, V_1^{\pi},\cdots,V_I^{\pi}\right)$ ~\cite{roijers2013survey}, where $f(\cdot,\cdot)$ is a nonlinear function. Hence, a multi-objective optimization problem can be converted to a single-objective optimization problem.

These value functions map a multi-dimensional policy value to a scalar according to the preferred policy on decision making and preference elicitation. Since all objectives are desirable, $V_{w}^{\pi}$ is monotonically increasing in all objectives. Given this monotonicity property, the solution set is a Pareto front, i.e., convex coverage set (CCS), that contains for any allowed policy $\pi^{\prime}$ with value $V^{\prime}$, any policy that has a greater or equal value in all objectives~\cite{roijers2013survey}. CCS can quantify the relationship between different policies subject to a number of objectives because one policy can yield good performance for one objective while poor performance for another objective. For instance, for objective $i_{1}$, solution $\theta$ is better when the loss $\mathcal{\hat{L}}^{i_{1}}\left ( \theta_{V_{i_{1}}}, \theta_{V_{j_{1}}}\right ) = - V_{w}^{\pi} < \mathcal{\hat{L}}^{i_{1}}\left ( \overline{\theta}_{V_{i_{1}}}, \overline{\theta}_{V_{j_{1}}}\right )$, while for objective $i_{2}$, solution $\overline{\theta}$ is better when $\mathcal{\hat{L}}^{i_{2}}\left ( \theta_{V_{i_{2}}}, \theta_{V_{j_{2}}}\right )>  \mathcal{\hat{L}}^{i_{2}}\left ( \overline{\theta}_{V_{i_{2}}}, \overline{\theta}_{V_{j_{2}}}\right )$. CCS defines the set of vector objective state values that the optimal value must reside in because every other vector objective state value not in the set will not be the best choice since there exists at least one vector in CCS that is not smaller than it. 

A formal definition of the convex coverage set is given below.
\begin{definition}
The convex coverage set, denoted as $CCS$, is the set of all actions and associated payoff values that are optimal for some $w$ of the scalarization function $f\left(w, V^{\pi}\right)$:
\begin{equation}
\exists V \in CCS: \forall V^{\prime}\in \pi: w^TV^{\prime}\leq w^TV,\forall w \in \mathbb{R}^d,
\end{equation}
where $u_w(\cdot)$ is the value when taking action $a$ based on the weight $w$.
\end{definition}

\subsection{Multi-objective Reinforcement Learning (MORL)}

We first introduce a few definitions that are needed in solving multi-objective optimization problems using (deep) reinforcement learning. Let a trajectory $\tau^{i}=\left\{q\left(x_{1}\right),q\left(x_{t+1}|x_{t},a_{t}\right),H\right\}$ consist of a distribution over initial observations $q\left(x_{1}\right)$ with a transition distribution $q\left(x_{t+1}|x_{t},a_{t}\right)$ and an episode length $H$. We define the loss $\mathcal{L}\left(x_{1},a_{1},...,x_{H},a_{H}\right)$ as the negated expected accumulated reward for a series of state-action pairs with length $H$ given by
\begin{equation}\label{eq:loss}
\mathcal{L}^{i}\left(f_{\theta_{\pi}}\right)=-E_{x_{t},a_{t}\sim f_{\theta_{\pi},\tau^{i}}}\sum_{t=1}^{H}R_{i}\left(x_{t},a_{t}\right),i=1,\cdots,I
\end{equation}
where $E$ is the expectation operation, $f_{\theta_{\pi},\tau_{i}}$ is the action distribution function determined by the policy $\pi$ that is assumed to be constructed using a neural network with $\theta_{\pi}$ acting as the weights. Another set of hyperparameter is needed to quantify the map from $V_{i}^{\pi}$ to $\mathcal{Y}^{i}$ defined as
\begin{equation}\label{eq:hyper_V_Y}
f\left(V_{i}^{\pi};w_{ii},w_{ij}\right): V_{i}^{\pi} \rightarrow \mathcal{Y}^{i},
\end{equation}
where $w_{ii}$ and $w_{ij}$ are the weights in the $i$th row of the IORM $W$.

\subsection{$W$ Update}\label{subsec:wab}

%(which are stated by there is a marginal weight of $ V_{S}^{*}\left ( w \right) $

We now provide a detailed description of how $W$ is updated within a batch. First, let's define the map from $\mathcal{X} $ to $\left \{ \mathcal{Y}^{i} \right \}_{i\in \left | I \right |}$ in a parametric form as $f^{i}\left(x;\theta_{V_{i}},\theta_{V_{j}}\right) : \mathcal{X} \rightarrow \left \{ \mathcal{Y}^{i} \right \}_{i\in \left | I \right |}$, where $\theta_{V_i}$ and $\theta_{V_j}$ are the aggregated weights of weights $\theta_{\pi}$ and hyperparameters $w_{ij}$. The main idea to update $W$ is to first obtain the CCS, then evaluate the marginal weights %that maximizes \textcolor{red}{unclear... why [w] in US while () in S below?}
%$$\Delta(w) =  V_{US}(w) - V_{S}^{*}\left ( w \right )$$ 
on the CCS, and finally use the best marginal weight to update $W$.%, where $\Delta(w)$ defines the maximum 

Because it is difficult to obtain the CCS directly, we employ the approximate optimistic linear support (AOLS) approach \cite{roijers2014bounded} to get an approximated set. The AOLS is a method that can gradually improve the approximation of the CCS. Given a maximum improvement threshold $\varepsilon>0$, the AOLS algorithm can compute an approximated $\varepsilon$-optimal set, denoted as $\overline{CCS}$, which may diverge from the optimal undominated set by at most $\varepsilon$. Consequently, its \textit{marginal weight} can be obtained. %which are shown in Algorithm 1.
Before a complete undominated set is obtained, a partial CCS can be obtained by evaluating the largest improvement for weights via the priority queue of the marginal weight in this step.
An element in the vector value function over a partial CCS is defined by %\textcolor{red}{what is $V_{\phi_{k}}^{i}$? what is $\phi_{k}$?}
$
V_{S}^{*}\left ( w \right )=\max_{V^{\pi}\in S}\, w\cdot V(s,\phi_V),
$
where $S$ is the partial CCS, $V(s,\phi_V)$ is the approximated objective state value vector based on the current critic networks using the current weights $\phi_V$, and $s$ is the current state.

%Similarly, the set of maximizing joint actions is $A_{S}\left ( w \right )=\underset{V_{\phi_{k}}^{i}\in S}{argmax}\, w\cdot V_{\phi_{k}}^{i}$.
AOLS always selects the marginal weight $w$ that maximizes an optimistic upper bound on the difference between $V_{\overline{CCS}}\left ( w \right )$ and $ V_{S}^{*}\left ( w \right )$, \textit{i.e.}, $V_{\overline{CCS}}\left (w \right )- V_{S}^{*}\left ( w \right )$, which can be updated iteratively to obtain a more accurate $\max_{w}  V_{CCS}(w)$. The pseudocode for AOLS is shown in the Algorithm~\ref{alg:DOL}. %the following  \algorithmref{alg:DOL}.

\begin{algorithm}[htbp]
\KwData{MOMDP: m, improvement threshold: $\varepsilon$}
\KwResult{CCS, $\Delta_{max}$}
\emph{S: empty partial CCS, W: empty list of explored marginal weight, Q: an empty priority queue of the initial marginal weight, $\Delta_{max}$: improvement}\\
\ForAll{extreme weights of infinite priority $w_{\max} =e_{1}$}
 {Q.add $(w_{\max},\infty)$}
\While{$\neg$ Q.isEmpty()  $\wedge \neg $ timeOut}{
$w_{i}^{j} \leftarrow$Q.pop()\\
$WV_{old} = WV_{old}  \cup  \left \{ \left ( w_{i}^{j} , w_{i}^{j}\cdot V(s,\phi_V) \right)\right \}$\\
\If{$ V(s,\phi_V) \notin S $}{
$S\leftarrow S \cup \left \{ V(s,\phi_V)  \right \}$\\
$W\leftarrow$ recompute marginal weight $V_{S}^{*}\left ( w \right )$\\
\For{$K\in 1,...,len(w)$}{
\If{$e_{K}\neq W$}{return$\left(e_{K},\infty \right)$}}
$V_{US}\left [ \cdot  \right ]\leftarrow$ $\forall$ weights in $W\left [ \cdot \right ]$, compute: max $w_{i}^{j}[K]\cdot v(s,\phi_V)$\\
subject to: $\forall \left (w_{i}^{j}[K],u \right )\in WV: w_{i}^{j}[K]\cdot v(s,\phi_V)\leq u+\varepsilon$\\$K\leftarrow \arg\max_{K}  V_{US}\left [ K \right ] - V_{S}^{*}\left ( W\left[K\right] \right )$\\
\If{$ V_{US}\left [ K \right ] - V_{S}^{*}\left ( W\left[K\right] \right ) > \varepsilon $}
{ Q.add$(W[K], V_{US}\left [ K \right ] - V_{S}^{*}\left ( W\left[K\right] \right ))$}}
$W\leftarrow W\cup \left \{ W\left [ K \right ] \right \}$
}
\caption{function AOLS$\left( m, \varepsilon,  V(s,\phi_k)\right)$}
\label{alg:DOL}
\end{algorithm}

\subsection{Value-function and Policy update}~\label{sec:VPU}
To obtain policy network and value function approximation network, we propose to adopt an actor-critic network with \textit{one actor network} and \textit{$I$ critic networks}, where the actor network is used to maximize the objective state value and each critic network is used to map from the state action pair to $\mathcal{Y}_{i}^{\pi}$. Assume that the actor network with weights $\theta_\pi$ generates actions via
$a = \pi \left(s; \theta_{\pi}\right).$
The weights $\theta_{\pi}$ can be updated using policy gradient given by~\cite{vamvoudakis2010online}:
\begin{align*}
\Delta \theta_{\pi}&\sim\sum_{k}\bigtriangledown_{\theta _{\pi}}\log \pi_{\theta}\left ( s_{k},a_{k} \right ) \delta_{k,t},
\end{align*}
where $\delta_{k,t}$ is the expected value of the $i$th objective, also known as the temporal difference (TD) residual of $\widehat{V}^\pi_{i}$ with discount $\gamma$~\cite{sutton2018reinforcement}, given by
\begin{align}\label{eq:aa}
\delta_{k,t} = &r_{i}\left ( s_{k,t},a_{k,t} \right )+\gamma\widehat{V}_{i}^{\pi(\theta_\pi)}\left ( s_{k,t+1};\phi_{V_{i}} \right )\\\notag&-\widehat{V}_{i}^{\pi \left( \theta^{-}_{\pi} \right)}\left ( s_{k};\phi^{-}_{V_{i}} \right )
\end{align}
where $r_{i}\left ( s_{k,t},a_{k,t} \right )$ is the immediate reward at the $t$th time step on the $k$th experience, $\widehat{V}_{i}^{\pi \left( \theta^{-}_{\pi} \right)}\left ( s_{k,t};\phi^{-}_{V_{i}} \right )$ is the approximation of the value function $V_i$ based on the old weights $\theta^{-}_{\pi}$ for the actor network and the old weights $\phi^{-}_{V_{i}}$ for the $i$th critic network, and $\widehat{V}_{i}^{\pi(\theta_\pi)}\left ( s_{k+1};\phi_{V_{i}} \right )$ is the approximation of the value function $V_i$ based on the updated weights $\theta_{\pi}$ for the actor network and the updated weights $\phi_{V_{i}}$ for the $i$th critic network.

%$V_{i}^{\pi}\left(s\right)$ is one element of the vector value function, $V_{i}\left(s;\phi_{V_{i}}\right)$ is the critic function.

For the $I$ critic networks, its $i$th neural network with hyperparameter $\phi_{V_{i}}$ is used to approximate each element in the vector value function  $V_{i}^{\pi}\left(s\right)$. Assume that the critic function is given by $V_{i}\left(s;\phi_{V_{i}}\right)$ with $\phi_{V_{i}}$ serving as the weights. The weights can be updated via
$
\Delta _{k}\phi _{V_{i}} \sim -\triangledown_{\phi _{V_{i}}} \sum_{k} \delta^2_{k,t}.
$

In the standard TD-residual method, the value of one action evaluated via \eqref{eq:aa} is an incremental form of value iteration. The key drawback of the standard TD-residual method includes the need for a large number of samples and large variance of policy gradient estimate. To address these issues, an existing approach, called generalized advantage estimator (GAE)~\cite{schulman2015high}, can be used to evaluate the action advantages and perform the policy updates using proximal policy optimization~\cite{schulman2017proximal,schulman2015trust,rockafellar1991scenarios}. The GAE is defined by:
\begin{align*}
&\widehat{A}_{t}^{GAE\left(\gamma ,\lambda \right)} \\
%=&\lim\limits_{H\to\infty} (1-\lambda)\sum_{j=1}^H \widehat{A}_{t}^{\left(j\right)}\\
=&\lim\limits_{H\to\infty} ( 1-\lambda ) \sum_{j=1}^H \lambda^{j-1} \sum_{k=1}^j \gamma^{j-1}\delta_{k,t+j-1} \\
=&\sum_{l=0}^{H}\left(\gamma \lambda \right)^{l}\delta_{k,t+l},
\end{align*}
where $\lambda\in \left[0,1\right]$ and $\gamma\in \left[0,1\right]$ adjusts the bias-variance tradeoff of GAE. 
 
After new weights of the advantage actor-critic network models are obtained, $V^{\pi}$ can be obtained via new samples using the updated policy. Afterwards, the procedure in Subsection~\ref{subsec:wab} can be implemented to obtain the updated $W$. The entire process will iterate until $V_{\overline{CCS}}\left ( w \right )- V_{S}^{*}\left ( w \right )<\epsilon$, where $\epsilon$ is a small threshold selected by users. 

\subsection{Explainable Planning Representation}~\label{sec:EPR}
To address the quantifiable inter-objective relationship in our algorithm, we adopt an explainable planning representation that enables automatic explanation of the planning rationale.
\subsubsection{Vocabulary for Quality Attributes (QA)} We map QA analytic models to domain-specific vocabulary to be used to generate verbal explanation. The vocabulary includes ``QA type'', ``optimization objective'', and ``QA property'' for the description of standard QAs.

\subsubsection{QA Language Templates} To generate verbal explanation of the objectives and the QA properties of a solution policy $\pi^\star$, we use predefined natural-language templates. Table~\ref{table_example} shows an example of verbal explanation of QA objectives and properties.

\begin{table}[htbp]
\caption{Verbal explanation of QA objectives and properties}
\centering
\scalebox{0.78}{
\begin{tabular}{c|c|c}
\hline
QA Type & Optimization Objective & QA Property\\
\hline
Standard measurement & ``maximize the alive bonus" & ``the expected alive bonus is 150"\\
\hline
\end{tabular}}
\label{table_example}
\end{table}

\subsubsection{Obtaining Alternative Policies}

Algorithm~\ref{alg:preI} outlines an approach for sampling alternative policies around the current policy. The key idea of the approach is to start with the QA values of the current solution policy $\pi: V_1^\pi(s),\cdots,V_I^\pi(s)$. For each QA $i$, we determine a new value $V_i^{\prime}$ that is more preferable than $V_i^\pi(s)$. Then, we construct a new planning problem with $I-1$ optimization objectives (namely, excluding the objective associated with the QA $i$), resulting in a new multi-objective value function subject to the constraint that the QA $i$ must be at least as good as $V_i^{\prime}$. Next, we select an optimal, constraint-satisfying solution value under $\pi^{\prime}$ for the new planning problem. The new policy (respectively, state value) provides an alternative of the current policy (respectively, state value associated with the current policy). This procedure will be executed iteratively until we obtain up to $M_i$ number of alternative policies for each $i$. The pseudocode of the algorithm is given below.

\begin{algorithm}[htbp]
\KwData{current policy $\pi$, state $s$, $V_1^\pi(s),\cdots,V_I^\pi(s)$, all $n-1$ attribute value functions $V_{\setminus 1}^\pi(s),\cdots,V_{\setminus I}^\pi(s)$, increment sizes of values $\Delta V_1,\cdots, \Delta V_I$, maximum values $M_{V_1},\cdots, M_{V_I}$, maximum number of alternatives $M_1,\cdots, M_I$}
\KwResult{A set of alternatives $\Pi^{\prime}$}
$\Pi^{\prime}\leftarrow \o $\\
$D \leftarrow$ attributes to be explored, e.g., $\left \{1,\cdots,n \right \};$ \\
\While{$D \neq \o$}{
$i \leftarrow$ remove an attribute from $D$\\
$count_i \leftarrow 0$\\
$V_i \leftarrow V_i^\pi(s)$\\
$V_{\setminus i} \leftarrow n-1$ attribute value function on all $V_{\setminus i}^\pi(s), j \neq i$\\
\While{$V_i \leq (M_{V_i}-\Delta V_i) \wedge count_i \leq M_i $}{
$V_i \leftarrow V_i + \Delta V_i$\\
$\pi: \overline{V}_i^\pi(s) \leftarrow$ all $\pi: V_j^\pi(s)$, where $j \neq i$\\
$V^{\prime} \leftarrow \underset{V}{\arg\max} \overline{V}_{\setminus i}^\pi(s)$, subject to $V_{\setminus i}^\pi(s) \geq V_i(s)$\\
\If{$V^{\prime}$ exists}{
$\Pi \leftarrow \Pi^{\prime} \cup \left \{V^{\prime}  \right \} $\\
$count_i \leftarrow count_i +1 $\\
\For{$j \neq i$}{
\If{$V_{j}^{\prime \pi}(s) \geq V_{j}^\pi(s)+\Delta V_j$}{$D \leftarrow D - \left \{j \right \}$}}}}
}
\caption{Pseudocode for the calculation of alternative multi-objective values (function AV2f $(\pi, \mathbf{V}^\pi(s), s, \Delta V, M_V, M_i )$)}
\label{alg:preI}
\end{algorithm}

\subsubsection{Semantic Explanation of Value Tradeoffs} Our value justification indicates the amount of gain-loss in the QAs if one were to choose each alternative value under the current policy. It then indicates preference towards the current policy by arguing that such gain is not worth the loss, reflecting the QA utility models underlying the multi-objective value function. We use a predefined natural language template for generating verbal justification: ``I could [improve these QAs to these values], by [carrying out this alternative policy] instead. However, this would [worsen these other QAs to these values]. I decided not to do that because [the improvement in these QAs] is not worth [the deterioration in these other QAs]".

\subsection{Overall Algorithm}
The pseudocode for the proposed V2f-MORL approach described in Subsections~\ref{subsec:wab},~\ref{sec:VPU}, and~\ref{sec:EPR} is given in the Algorithm~\ref{alg:seq}.
\begin{algorithm}[htbp]
\KwData{initial policy parameters $\theta_{0}$, initial value function parameters $\phi_{0}^{V_{i}}$, initial vectorized weights based on each objective $w_{ij}$, $CCS$}
\KwResult{PPO\_Model}
\For{$i=1$ to $I$}{
\For{$k=0,1,2,...,K$}
{ Collect set of trajectories $D_{k}^{i}=\left \{ \tau _{n}^{i} \right \}$ by running policy $\pi_{k}=\pi\left ( \theta_{k} \right )$ in the environment\\
Compute rewards-to-go $\hat{R_{t}}$\\
Update $ V(s,\phi_k)$\\
 Compute $w_{i\left [ \cdot \right ] } $ by function AOLS$\left( m, \varepsilon, V(s,\phi_k) \right)$\\
 $V_{i}^{\phi_{k}} = w_{i\left [ \cdot \right ] } \times  V(s,\phi_k) $\\
Compute advantage estimates $\hat{A_{t}^{i}}$ using GAE method based on the current value function $V_{i}^{\phi_{k}}$\\
 Update the policy by maximizing the PPO-Clip objective:
\begin{align*}
\theta_{k+1} = &\underset{\theta}{\arg\max}\frac{1}{\left |D _{k}^{i} \right | T}\sum_{\tau_{n}^{i}\in D _{k}^{i} } \sum_{t=0}^{T}\min\Big( \frac{\pi_{\theta}\left ( a_{t}|s_{t} \right )}{\pi_{\theta_{k}}\left ( a_{t}|s_{t} \right )} \\
& A^{i,\pi_{\theta_{k}}}\left( s_{t},a_{t} \right ), g\left ( \epsilon, A^{i,\pi_{\theta_{k}}}\left( s_{t},a_{t} \right )  \right ) \Big)
\end{align*}
via stochastic gradient ascent (e.g., Adam)\\
Fit value function by regression on mean-squared error:
\begin{align*}
\phi_{k+1}^{V_{i}}=\underset{\phi_{k}^{V_{i}}}{\arg\min}\frac{1}{\left | D_{k}^{i} \right |T}\sum\limits_{\tau_{n}^{i}\in D _{k}^{i} } \sum\limits_{t=0}^{T}\left ( V_{i}^{\phi_{k}} \left ( s_{t} \right )-\hat{R_{t}^{i}} \right )^{2}
\end{align*} via stochastic gradient descent.\\
$\Pi \leftarrow$ function AV2f $(\pi, \mathbf{V}^\pi(s), s, \Delta V, M_V, M_i )$}
}
\caption{Pseudocode for the calculation of vectorized multi-objective values (V2f-MORL)}
\label{alg:seq}
\end{algorithm}
\section{EXPERIMENTS}\label{sec:exp}
\subsection{Setup}
We here select the testing environment Ant-v2 on the MuJoCo physics engine~\cite{todorov2012mujoco}, and select four objectives: Reward Control (Rctrl), Reward Contact (Rcont), Reward Survive (Rsurv), and Reward Forward (Rfor). We use the proximal policy optimization clipping algorithm with $\epsilon=0.2$ as the optimizer. The discounting factor is selected as $\gamma=0.99$. One episode, characterizing the number of time steps of the vectorized environment per update, is chosen as $2048$. For stabilization purposes, we execute parallel episodes in one batch. The batch size is chosen as the product of the episode size and the number of environment copies simulated in parallel. The number of environment copies is selected as $8$. The parameters are optimized using the Adam algorithm~\cite{kingma2014adam} and a learning rate of $3\times10^{-4}$. All of the experiments were performed using TensorFlow, which allows for automatic differentiation through the gradient updates~\cite{abadi2016tensorflow}. %The result takes the average over experiments.

\begin{figure}[thpb]
      \centering
\includegraphics[width=8cm]{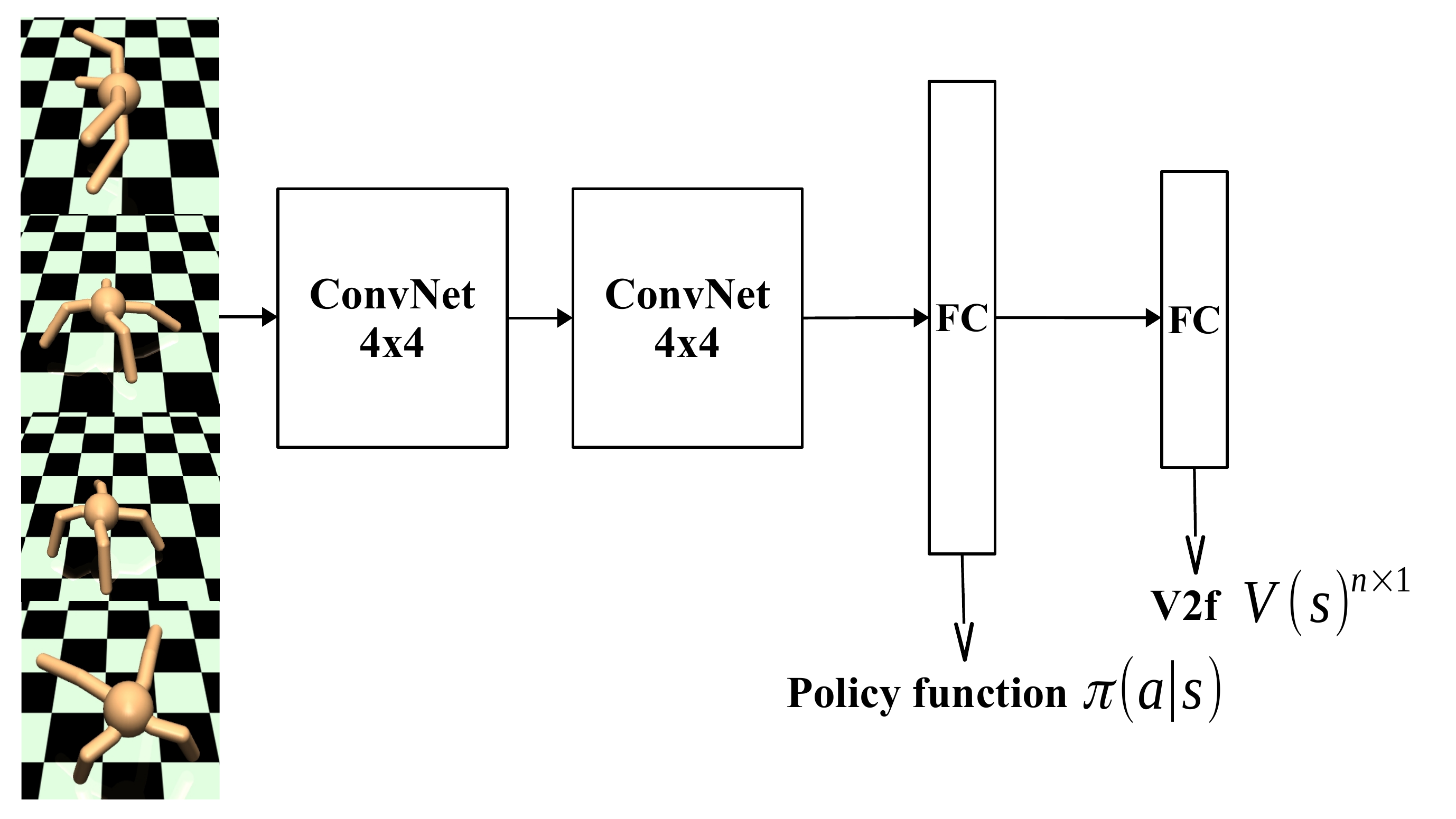}
      \caption{The architecture of the shared parameter network}
      \label{figure:ar}
   \end{figure}

The parametric form $f^{i}\left(x;\theta_{V_{i}},\theta_{V_{j}}\right) : \mathcal{X} \rightarrow \left \{ \mathcal{Y}^{i} \right \}_{i\in \left | I \right |}$, where $\theta_{V_{i}}$ is the objective specific weight and $\theta_{V_{j}}$ is the inter-objective weight, is a CNN whose structure is shown in Fig.~\ref{figure:ar}. The architecture specification is given in Table~\ref{table_cnn}.  The FC8 and FC4 correspond to the 8-action policy $\pi(\cdot|s_t)$ and the value function $\mathbf{V}(s_t)$. In the experiments, the screen is resized to an $84\times 84\times3$ RGB image as the network input.

\begin{table}[h]
\caption{CNN architecture}

\centering
\scalebox{1}{\begin{tabular}{c|c|c|c|c}
\hline
Layer $\#$ & 1 & 2 & 3 & 4\\
\hline
Parameters & $C4\times 4-32S2$ & $C4\times 4-32S2$ & $FC8$ & $FC4$\\
\hline
\end{tabular}}
\label{table_cnn}
\end{table}

\subsection{Accuracy vs Episodes}\label{AE}

\begin{figure}[htbp]
\begin{center}
\includegraphics[width=7cm]{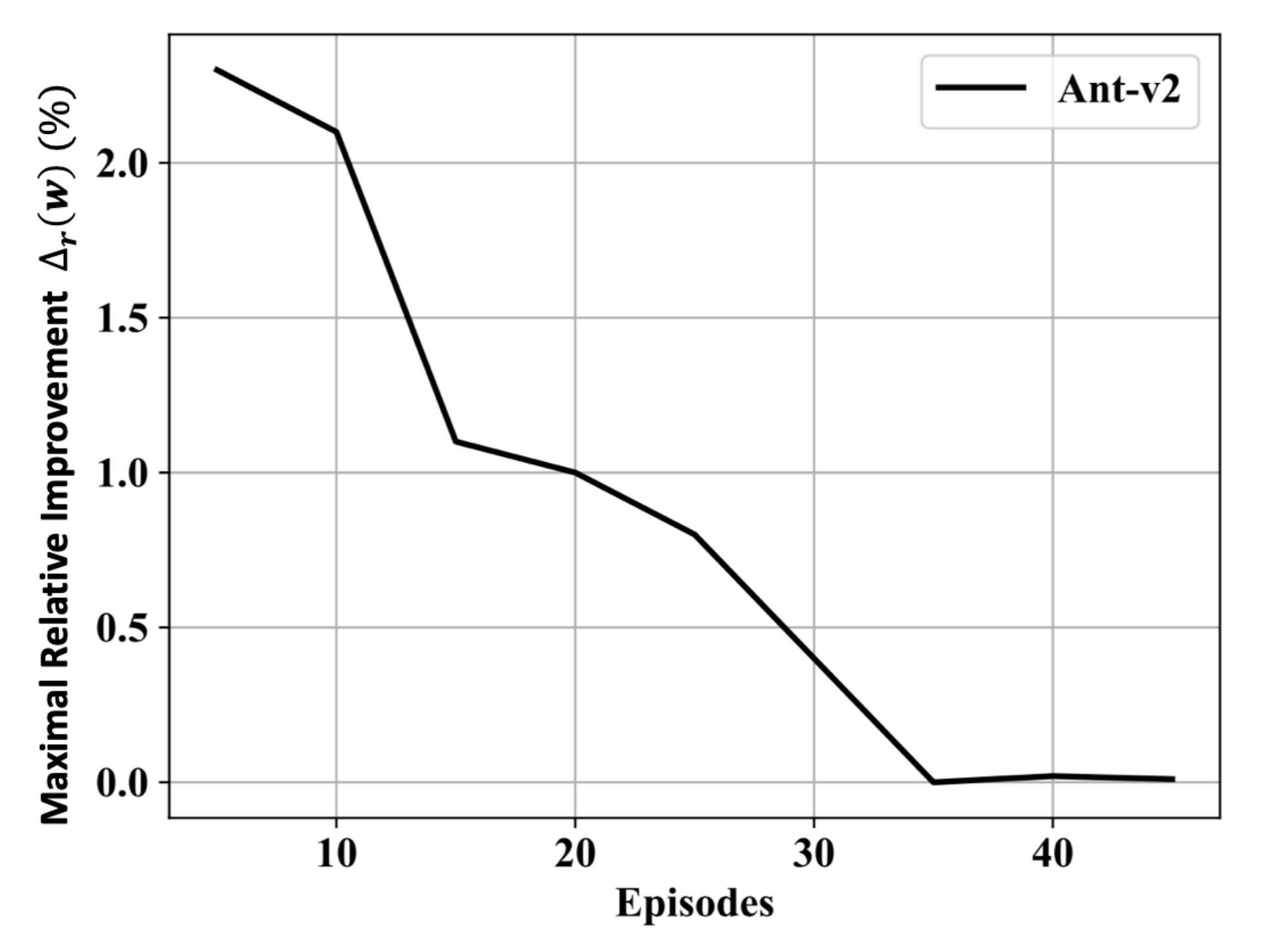}
\caption{Evolution of $\Delta_{r}\left(w\right)$ with respect to episodes in percentage}
\label{fig:Evol}
\end{center}
\end{figure}

%\textcolor{red}{Fig. 4 uses "batches" as x-axis while the description here uses episodes. inconsistent}
We further investigated the effects of the number of training episodes (including series of time steps) on the maximal relative improvement of the CCS. Fig.~\ref{fig:Evol} shows how the maximal relative improvement $\Delta_{r}\left(w\right)=\frac{V_{\overline{CCS}}(w) - V_{S}^{*}(w)}{V_{\overline{CCS}}\left(w\right)}$ of the CCS evolves with respect to the number of episodes. It can be seen from Fig.~\ref{fig:Evol} that the error is highly affected by the number of training episodes. Although the proposed method is unable to provide sufficient accuracy to build the CCS initially, the deviation will gradually decrease to $0$ when the number of episodes is $40$. %We can also observe from Fig. 4 that the networks are not overfitting beyond that because the maximal relative improvement stays at $0$.

\subsection{Testing Results and Discussion}\label{Test}
For the simplicity of presentation, we call our proposed method PPO\_CCS method. To show the benefit of the proposed PPO\_CCS method, we show the results when (1) multi-objective optimization is solved via one single-objective optimization, and (2) the marginal weight in our method is replaced by the corner points of the CCS. All these results are based on the MuJoCo simulator~\cite{todorov2012mujoco}.

Our goal is to make a four-legged ant-v2 walk forward as fast as possible while saving cost simultaneously. More specifically, our goal is to maximize the reward forward and the reward survive while minimizing the reward control and the reward contact. We take the current reward function in the OpenAI Gym environments as a baseline, use the cumulative reward trained by the single objective PPO as a benchmark \cite{brockman2016openai,dhariwal2017openai}, and compare it with our proposed method.

Table~\ref{mot1} shows the rewards using PPO\_CCS and PPO with single objective. It can be observed that the proposed PPO\_CCS yields higher reward. It can also be observed that PPO\_CCS can generate higher rewards in most cases because PPO\_CCS can optimize multiple objectives simultaneously, while the PPO with single objective does not seek to optimize multiple objectives. 

\begin{table}
\centering
% space before first and after last column: 1.5pc
% space between columns: 3.0pc (twice the above)
\setlength{\tabcolsep}{1pc}
% -----------------------------------------------------
% adapted from TeX book, p. 241
\newlength{\digitwidth} \settowidth{\digitwidth}{\rm 0}
\catcode`?=\active \def?{\kern\digitwidth}
% -----------------------------------------------------
\caption{Multi-objective Value for Ant-v2}
\label{mot1}
\label{tab:effluents}
\scalebox{0.75}{
%\hspace{-2cm}
\begin{tabular*}{\textwidth}{@{}l@{\extracolsep{\fill}}rrrr}
\cline{1-4}
                 & \multicolumn{2}{l}{~~~~~~~~~~~~~~~~~~Single-objective}
                 &\multicolumn{2}{l}{~~~Multi-objective} \\
\cline{2-3} \cline{3-4}
                 & \multicolumn{1}{r}{Rfor}
                 & \multicolumn{1}{r}{Rsurv}
                 & \multicolumn{1}{r}{PPO\_CCS}
       \\
\cline{1-4}
Rsurv    & $ 76.341 \pm 18.566 $ & $77.445\pm14.542$ & $  92.546\pm18.446$   \\
Rfor & $ 0.541 \pm 0.026$ & $0.740 \pm 0.057$ &  $0.818\pm0.147$     \\
Rctrl      & $-4.011\pm0.730$ & $-4.019\pm0.825$ & $ -5.025\pm0.011$  \\
Rcont          & $ -3.442\pm0.250$ & $-3.596\pm0.011$ &   $-4.\pm0.023$       \\
\cline{1-4}
\multicolumn{5}{@{}p{140mm}}{Average and standard deviation $\left(\mu\pm\sigma\right)$ of multi-objective values.}
\end{tabular*}}
\end{table}

\subsection{Natural Language Representation Demonstration}

We demonstrate three semantic representations that are generated under different policies using the proposed algorithm in Algorithm~\ref{alg:preI}. The result is shown in Fig.~\ref{fig:repr}. The first representation provides verbal explanation of the state values under the selected policy while the second and third representations provide verbal explanations why alternative policies were not selected. This is because the algorithm: (i) searched to reduce Rctrl on the CCS and found a different multi-objective value function that increases Rcont, decreases Rsurv, and decreases Rfor, and (ii) searched to reduce Rcont and found a different multi-objective value function that increases Rctrl, decreases Rsurv, and decreases Rfor.

   \begin{figure}[hhhh]
      \centering
      \vspace{.2cm}
      \framebox{\parbox{3in}{``\textit{I aim to maximize the reward forward and the reward survive while minimizing the reward control and the reward contact. I plan to move forward. The Rctrl is} -5.025, \textit{Rcont is} -4, \textit{Rsurv is} 92.546, \textit{and Rfor is} 0.818."}}
            \framebox{\parbox{3in}{``\textit{I could decrease the} Rctrl \textit{to} -8.236, \textit{by move forward in another set of actions instead. However, this would decrease the} Rcont \textit{by} -1.953, \textit{decrease the} Rsurv \textit{by} 45.045, \textit{and decrease the} Rfor \textit{by} 0.417. \textit{I decided not to do that because the decrease in the} Rctrl \textit{is not worth the increase of the} Rcont, \textit{the decrease of the} Rsurv, \textit{and the decrease of the} Rfor." }}
                  \framebox{\parbox{3in}{``\textit{I could also decrease the} Rcont \textit{to} -4.081, \textit{by move forward in another set of actions instead. However, this would decrease the} Rctrl \textit{by} -0.031, \textit{decrease the} Rsurv \textit{by} 7.882, \textit{and decrease the} Rfor \textit{by} 0.174. \textit{I decided not to do that because the decrease in the} Rcont \textit{is not worth the increase of the} Rctrl, \textit{the decrease of the} Rsurv, \textit{and the decrease of the} Rfor. "}}
    \caption{Natural language representation}
    \label{fig:repr}
   \end{figure}

\section{CONCLUSIONS}

In multi-objective optimization problems, the possibly conflicting objectives necessitates a trade-off when multiple objectives need to optimize simultaneously. A typical approach is to minimize a loss of weighted linear summation of all objective functions. However, this approach can hardly guarantee good performance on individual objectives because it is very difficult to determine the right weights due to the lack of knowledge in inter-objective relationships. To address the challenge, we proposed a vector value function based multi-objective deep reinforcement learning to solve high-dimensional multi-objective decision making problems. The proposed method optimizes vectorized proxy objectives sequentially based on proximal policy optimization, actor-critical network, and the derivation of optimal weights via marginal weight. 
%The Proximal Policy Optimization (PPO) is used as our standard RL algorithm for the actor network.

By explicitly quantifying inter-objective relationship via relationship matrix, the relative importance of the objectives unknown \textit{a prior} can be obtained via reinforcement learning. Each entry in the relationship matrix specifies and explains the relative impact of one objective on another objective in the optimization step. Moreover, in order to address the interpretability of the proposed V2f-MORL approach, we proposed a new approach to generate alternative multi-objective values/policies to automatically explain the rationale behind decided actions/policies.

%There remain many interesting directions for future research into multi-objective optimization. For example, how to evaluate the impact of a perturbation on a row of $W$ on individual objective $V_{i}^{\pi}$? How to consider constraints on selected objectives?  


%%%%%%%%%%%%%%%%%%%%%%%%%%%%%%%%%%%%%%%%%%%%%%%%%%%%%%%%%%%%%%%%%%%%%%%%%%%%%%%%

%%%%%%%%%%%%%%%%%%%%%%%%%%%%%%%%%%%%%%%%%%%%%%%%%%%%%%%%%%%%%%%%%%%%%%%%%%%%%%%%

%%%%%%%%%%%%%%%%%%%%%%%%%%%%%%%%%%%%%%%%%%%%%%%%%%%%%%%%%%%%%%%%%%%%%%%%%%%%%%%%

\bibliographystyle{IEEEtran}
\bibliography{references1}

\end{document}